\documentclass[amsmath,superscriptaddress,showpacs,twocolumn,aps,prl]{revtex4}
\usepackage{bm}
\usepackage{graphics}
\usepackage{graphicx}

\newcommand{\bq}{\begin{equation}}
\newcommand{\ee}{\end{equation}}

\begin{document}
\title{Spin-Hall edge spin polarization in a ballistic 2D electron system}
\author{V. A. Zyuzin}
\affiliation{Department of Physics, University of Utah, Salt Lake
City, Utah 84112, USA}
\author{P. G. Silvestrov}
 \affiliation{Theoretische
Physik III, Ruhr-Universit{\"a}t Bochum, 44780 Bochum, Germany}

\author{E. G. Mishchenko}
\affiliation{Department of Physics, University of Utah, Salt Lake
City, Utah 84112, USA}

\begin{abstract}

Universal properties of spin-Hall effect in ballistic 2D electron
systems are addressed. The net spin polarization across the edge of
the conductor is second order, $\sim \lambda^2$, in spin-orbit
coupling constant independent of the form of the boundary potential,
with the contributions of normal and evanescent modes each being
$\sim \sqrt{\lambda}$ but of opposite signs. This general result is
confirmed by the analytical solution for a hard-wall boundary, which
also yields the detailed distribution of the local spin
polarization. The latter shows fast (Friedel) oscillations with the
spin-orbit coupling entering via the period of slow beatings only.
Long-wavelength contributions of evanescent and normal modes exactly
cancel each other in the spectral distribution of the local spin
density.

\end{abstract}

\maketitle

{\it \underline{Introduction.}} Spintronics addresses interplay of
spin and orbital degrees of freedom in various transport, optical,
etc.~phenomena with the ultimate goal of achieving  spin
manipulation in nanostructures. Special place in spintronics belongs
to the spin-Hall effect predicted a long time ago \cite{DP}, which
recently entered the era of experimental observation
\cite{exp1,exp2,exp3}. Spin-Hall effect is characterized by a
boundary (edge) spin polarization resulting  when electric current
is flowing through the system. It is customary classified into
``extrinsic'' (impurity-driven) \cite{H,Z,ERH,DS} and ``intrinsic''
(band-structure induced) \cite{MNZ,Sinova} types. Initially theories
of spin-Hall effect addressed such auxiliary quantity as spin
current (for the review see Refs.~\cite{reviews}) in infinite
systems, but later the emphasis shifted towards direct calculation
of spin polarization in confined geometries. For diffusive systems
the search is to complement the coupled spin-density diffusion
equations \cite{MSH,BNM} with suitable boundary conditions
\cite{Chu,AB,GBD,Blei,Tser,RGD}.

While it is now understood that in 2D systems spin-Hall effect
generally occurs with more complicated spin-orbit couplings, any
amount of disorder destroys spin-Hall effect in infinite systems
with linear coupling \cite{reviews}. It is, therefore, important to
establish whether pure ballistic systems (without disorder) can
exhibit non-zero spin-Hall polarization. Driven by this motivation,
studies of intrinsic spin-Hall effect in ballistic finite-size
systems had been initiated, mostly by means of numerical methods
\cite{NWS,nik}. It is significant  to realize that the edge spin
polarization in ballistic systems appears not as a result of
electric field-driven acceleration of electron momenta (and
associated with it precession of spins). As well known, electric
field is absent inside an ideal ballistic conductor connected to
reflectionless leads \cite{Datta}. Spin-Hall spin accumulation in
ballistic systems is due to the edge precession only. When the
populations of left-moving and right moving states are different,
the boundary scattering results in oscillatory (Friedel) edge
polarization which is perpendicular to both the electric current and
the normal direction to the boundary. Such polarization was
considered numerically in Refs.~\cite{usaj} for a 2D electron gas
(2DEG). The case of a 3D hole semiconductor has also been analyzed
recently \cite{SG}. A possibility of distinguishing edge effects
from spin transport has been addressed experimentally in
Ref.~\cite{diff}. Edge spin polarization in parabolic quantum wires
has been considered in Ref.~\cite{EEL}.

In the present paper we resolve analytically the boundary problem
for a ballistic 2D electron gas with linear spin-orbit coupling
\cite{BR} and calculate the non-equilibrium edge spin-polarization
in a wide strip connected to ideal leads with chemical potentials
shifted by the applied voltage. We present a general argument that
the out-of plane spin polarization integrated over the lateral
direction has a {\it universal} value, {\it independent} of the
particular shape of the confining boundary potential $U(x)$. In the
limit of weak spin-orbit coupling, $\lambda \ll v_F$,
\begin{equation}
\label{result} \int_{-\infty}^\infty s_z(x) dx = - \frac{\lambda^2
eV}{12\pi^2 v^3_F},
\end{equation}
where $eV$ is the difference of the chemical potentials in the two
leads, and $v_F$ is the bulk value of the Fermi velocity (which is
the same for both spin-split subbands).

We then illustrate how this result arises from microscopic
calculations in a model of a sharp boundary by obtaining the
electron Green's functions in a concise analytical form. The
obtained spin density is approximated by the expression ($\hbar=1$),
 \begin{equation}
 \label{localapprox}
s_z(x)\approx \frac{eV}{2\pi^2v_F x}\cos{(2mv_Fx)} \sin^2{(m\lambda
x)}.
 \end{equation}
It is remarkable that the spin-orbit coupling constant enters via
the period of beating only.
\begin{figure}[h]
\resizebox{.48\textwidth}{!}{\includegraphics{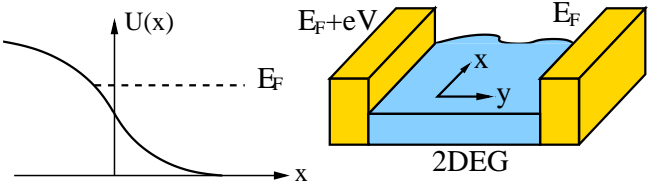}}
\caption{(Color online) Geometry of the system. Ideal leads filled
by equilibrium electrons up to the chemical potentials shifted by
the applied bias. The edge is formed by a confining potential $U(x)$
vanishing for $x\to \infty$.}
\end{figure}

{\it \underline{Net spin polarization.}} Consider a semi-infinite
ballistic 2DEG described by the Hamiltonian
\begin{eqnarray}
\label{Ham}
 H=\int d{\bf r} ~\hat \psi^\dagger \left[
-\frac{{\bm \partial}^2}{2m}-i\lambda (\hat{\sigma}_x
\partial_y-\hat{\sigma}_y \partial_x) +U(x) \right] \hat \psi ,
 \end{eqnarray}
where potential $U(x)$ ensures boundary confinement (see Fig.~1).
For the sake of simplicity we present derivation for the case of
'Rashba' spin-orbit interaction, though calculations  for the
'Dresselhaus' coupling  \cite{Dress} are completely analogous
\cite{both}. The system is attached to two ideal reflectionless
leads injecting equilibrium electrons into 2DEG. The chemical
potentials of the leads are shifted by the applied voltage, $eV$.

Since $k_y$ is an integral of motion (in case of reflectionless
leads), it is convenient to use the Fourier representation along the
$y$-axis for the electron operators, $\hat \psi({\bf
r})=\sum_{k_y}\hat c_{k_y}(x)e^{ik_yy}$. One can now derive the
equation of motion for the expectation value of the electron spin
operator, ${\bf s}(k_y,x)=\frac{1}{2} \langle \hat
c^\dagger_{k_y}(x) \hat {\bm \sigma} \hat c_{k_y}(x)\rangle$, which
can be readily written in the form,
\begin{equation}
\label{conserv}
\partial_t s_y(k_y,x)=- \partial_x
J^y_x(k_y,x)-2\lambda k_y s_z(k_y,x).
\end{equation}
Here $J_x^y$ stands for the conventional operator of spin-current,
i.e.,
$$ J^y_x(k_y,x)=\frac{i}{4m}\langle \nabla_x \hat c^\dagger_{k_y}
\hat\sigma_y \hat c_{k_y} -  \hat c^\dagger_{k_y} \hat \sigma_y
\nabla_x \hat c_{k_y}\rangle -\frac{\lambda}{2}\langle \hat
c^\dagger_{k_y} \hat c_{k_y} \rangle.
$$
In a steady state the lhs of Eq.~(\ref{conserv}) vanishes.
Integrating Eq.~(\ref{conserv}) over the $x$-direction, we obtain
for the net spin polarization,
\begin{equation}
\label{polar} \int_{-\infty}^\infty s_z(x) dx = -\frac{1}{2\lambda}
\sum_{k_y} \frac{1}{k_y} J^y_x(k_y,\infty).
\end{equation}
It is straightforward to calculate the value of the ($k_y$-resolved)
spin current $J^y_x(k_y,\infty)$ inside the bulk of a 2D system:
\begin{equation}
\label{spin_current} J^y_x(k_y,\infty)= -\frac{1}{2}\sum_{\beta=\pm
1} \sum_{k_x} \left(\lambda+\frac{\beta k_x^2}{mk} \right)
n_\beta(k_x,k_y),
\end{equation}
where $n_\beta(k_x,k_y)$ stands for the population of different
momentum states in the subband $\beta$. Only ``uncompensated''
states contribute to the non-equilibrium spin polarization given by
Eqs.~(\ref{polar}-\ref{spin_current}); these states describe
electrons that originate in the left lead ($k_y>0$) and belong to
the energy interval near the Fermi energy, $E_F<k^2/2m+\beta
k\lambda < E_F+eV$. The integral (\ref{polar}) diverges
logarithmically at $k_y \to 0$. Assuming the same infrared cut-off
in both subbands, $\widetilde k$, we observe that the diverging
$\ln{\widetilde k}$-contributions in the two subbands cancel each
other, yielding in the linear (in $V$) response,
\begin{equation}
\label{s_density} \int_{-\infty}^\infty s_z dx
=\frac{eV}{2\lambda(2\pi)^2} \left(\frac{2\lambda}{v_F} -
\ln{\frac{v_F+\lambda}{v_F-\lambda}}\right)
\end{equation}
where $v_F=\sqrt{2E_F/m+\lambda^2}$ is the Fermi velocity. Expanding
this {\it general} result to the lowest non-vanishing order in
$\lambda/v_F$ we recover the net boundary polarization, Eq.
(\ref{result}).



{\it \underline{Evanescent modes}}. The reflection at the boundary
mixes the two bulk subbands. Those states that belong  to the
domain, $k^+<k_y<k^-$, where $k^\pm =m(v_F\mp \lambda)$, refereed to
as evanescent states \cite{usaj}, are characterized by exponentially
decaying contribution from the upper ($+$) subband. Repeating the
calculations leading to Eq.~(\ref{s_density}) but now for the
evanescent domain only, we obtain,
\begin{equation}
\label{s_density_ev} \int_{-\infty}^\infty s_z^{\text{ev}} dx
=\frac{eV}{2\lambda(2\pi)^2} \left(2\sqrt{\frac{\lambda}{v_F}} -
\ln{\frac{1+\sqrt{\lambda/v_F}}{1-\sqrt{\lambda/v_F}}}\right).
\end{equation}
Remarkably, the net evanescent contribution is $\sim \sqrt{\lambda}$
and is {\it largely} canceled by the contribution from the normal
domain $k_y<k^+$, yielding Eq.~(\ref{result}) which is quadratic
in~$\lambda$. This cancelation occurs for local spin density as
well, see Eq.~(\ref{localapprox}).
\begin{figure}
\includegraphics[width=5.0cm,angle=-90]{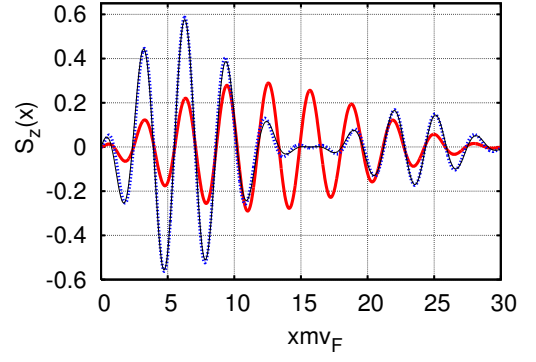}
\caption{(Color online) Dependence of the local spin polarization
(\ref{x}), in units of $eVm/8\pi^2$, on the distance to the boundary
for different values of spin-orbit coupling constant. Solid (red)
line: $\lambda/v_F=0.1$, dotted (blue) line: $\lambda/v_F=0.2$,
solid (black) line utilizes the approximate formula
(\ref{localapprox}) for $\lambda/v_F=0.2$. The plot of
Eq.~(\ref{approximate}) is indistinguishable from the exact
Eq.~(\ref{x}) on this scale. } \label{fig2}
\end{figure}

{\it \underline{Electron Green's function}}. Microscopic calculation
of the local spin polarization can be most simply performed with the
help of the electron Green's functions,
\begin{equation}
\label{micro} {\bf s}(x)=i\frac{eV}{4\pi} \text{Tr}
\int\limits_0^{k^-}\frac{dk_y}{2\pi}
[G^R_{k_yE_F}(x,x)-G^A_{k_yE_F}(x,x)]\hat{\bm \sigma},
\end{equation}
where $G^R_{k_yE}(x,x')$ is the retarded Green's function. Its
advanced counterpart satisfies the condition $G^A_{k_yE}(x,x')=G^{R
\dagger}_{k_yE}(x',x)$. The summation over energy in
Eq.~(\ref{micro}) is performed over the  narrow strip of width $eV$,
similar to Eqs.~(\ref{spin_current}-\ref{s_density}).

To illustrate how spin polarization arises from the solution of the
Schr\"odinger equation, let us solve the problem of a hard-wall
boundary: $U(x)=0$, for $x>0$ and $U(x)=\infty$ for $x<0$. The case
of a smooth boundary where electrons adiabatically follow
semiclassical trajectories for spin-split subbands \cite{SM} will be
considered separately \cite{ZSM}.

For a  plane wave, $\sim e^{ik_y y}$, the equation for
the Green's functions for ($x,x'>0$) is
\begin{equation}
\label{Green} \left[\frac{\partial^2_x}{2m}-\lambda
(i\hat\sigma_y\partial_x+\hat \sigma_x k_y )
 +E' \right]\hat G(x,x')=-\delta(x-x'),
\end{equation}
where the subscripts $k_y$ and $E$ are omitted for simplicity, and
$E'=E-{k_y^2}/{2m}$. The boundary condition for the impenetrable
wall is $G(x,0)=G(0,x')=0$. We solve the problem by first noting
that the following function $\hat{\cal L}(x)$ satisfies both the
homogeneous equation (\ref{Green}) and the boundary condition
$\hat{\cal L}(0)=0$,
\begin{equation}
\label{psi} \hat{\cal L}(x)  = \frac{1}{i\sum_{\beta} k^\beta}
\sum_\beta \frac{k^{\beta}}{k_{x}^{\beta }} \left(e^{ik_{x}^{\beta}
x} \hat B_\beta- e^{-ik_{x}^{\beta} x} \hat B_\beta^* \right),
\end{equation}
here $^*$ stands for the simple complex (not Hermitian) conjugate;
the sum is taken over both subbands, with the projection matrix for
the subband $\beta$  defined as
\begin{equation}
\label{beta}  B_\beta= \frac{1}{2} \left( 1 +\beta
\frac{k_y}{k^\beta}\hat \sigma_x - \beta
\frac{k_{x}^{\beta}}{k^{\beta} }\hat \sigma_y \right),
\end{equation}
 where the absolute value of the electron momentum $k^\beta$ is
defined above Eq.~(\ref{s_density_ev}) and its $x$-component
 is $k_{x}^{\beta}
=\sqrt{(k^{\beta})^2-k_y^2}$. Here we concentrate on the  normal
modes, where both $k_x^{\pm}$ are real; rather simple modifications
for the evanescent domain  (where $k_x^+$ is imaginary) are outlined
below.

Using the function (\ref{psi}) we can readily construct the solution
for the inhomogeneous equation (\ref{Green}) which satisfies the
boundary condition $G(0,x')=0$,
\begin{equation}
\label{gx}\hat G(x,x')  = - 2m[\hat{\cal L}(x) \hat
A(x')+\Theta(x-x')\hat {\cal L}(x-x')],
\end{equation}
where $\hat A(x')$ is yet unknown matrix. Since Green's function has
to obey both the boundary condition $G(x,0)=0$ and the equation
conjugated to Eq.~(\ref{Green}), the matrix $\hat A(x')$ must be a
homogeneous solution satisfying the condition $\hat A(0)= - 1$. This
determines it up to some constant matrix $\hat C$ different for the
retarded and advanced Green's functions, $ \hat A(x')= \hat C \hat
{\cal L}^\dagger(x') -\partial_{x'} \hat {\cal L}^\dagger(x')$,
\begin{eqnarray}
\label{gx1}\hat G_{R,A}(x,x')  &=& -2m \Bigl[ \hat{\cal L}(x) \hat
C_{R,A} \hat {\cal L}^\dagger(x')-\hat{\cal L}(x) \partial_{x'}
\hat{\cal L}^\dagger(x') \nonumber\\ && +\Theta(x-x')\hat {\cal
L}(x-x') \Bigr].
\end{eqnarray}
The constant $\hat C_{R}$ ($\hat C_{A}$) is most simply determined
from the condition that the retarded (advanced) Green's function
does not contain the waves $\sim e^{-ik_{x}^{\beta }x}$
($e^{ik_{x}^{\beta }x}$) propagating to (from) the boundary in the
region $x>x'$. The calculations are straightforward but rather
tedious. As a result one obtains,
\begin{eqnarray}
\label{c} \hat C_{R,A}=\mp \frac{i}{2}\left( k_x^++k_x^-\right) \mp
\frac{i}{2k_y} \left(k^-k_x^+-k^+k_x^- \right) \hat \sigma_x
\nonumber\\  -im\lambda \hat \sigma_y +\frac{1}{2k_y} \left(k^+k^-
-k_y^2-k_x^+k_x^- \right)\hat \sigma_z,
\end{eqnarray}
where the upper (lower) sign corresponds to $ \hat C_{R}(\hat
C_{A})$.


{\it \underline{Spin polarization}}. Making use of the derived
Green's function we can now calculate the local spin polarization
(\ref{micro}). With the help of Eqs.~(\ref{gx1}) and (\ref{c}) we
obtain
\begin{widetext}
\begin{eqnarray}
\label{x} s_{z}(x)=
\frac{eV}{2(2\pi)^2mv_F^2} \left\{ \int_0^{k^{+}} \frac{dk_y}{ k_y}~
(k^{+}k^{-}+k^{2}_{y}-k^{+}_{x}k^{-}_{x} )
\Big[\sin(2k^{+}_{x}x)+\sin(2k^{-}_{x}x)-2\sin((k^{+}_{x}
+k^{-}_{x})x) \Big] \right.\nonumber\\ \left.- \int_{k^{+}}^ {k^{-}}
\frac{dk_y}{ k_y}~\Big[k^{-}_{x}\kappa (e^{ik_x^-x}-e^{-\kappa x})^2
-2(k^+k^-+k_y^2+ik_x^-\kappa)\sin(k^{-}_{x}x)(\cos(k^{-}_{x}x)-e^{-\kappa
x}) \Big]\right\}
\end{eqnarray}
\end{widetext}
The first line here is the contribution of the normal modes while
the second line comes from the evanescent modes, where
$\kappa=\sqrt{k_y^2-(k^+)^2}$ \cite{evanesc}. By calculating the
integral over $x$ it is straightforward to verify that the net
contribution of the evanescent modes satisfies
Eq.~(\ref{s_density_ev}), being $\sim \sqrt{\lambda}$. This is
mostly canceled by the contribution from the normal modes. The total
net contribution of both the normal and evanescent states yields
Eq.~(\ref{s_density}) in agreement with our general argument based
on the equation of motion for spin operators, Eq.~(\ref{conserv}).
Behavior of the local spin density (\ref{x}) is shown in Fig.~2. In
the most relevant limit, $\lambda \ll v_F$, Eq.~(\ref{x}) can be
simplified to:
\begin{eqnarray}
\label{approximate} s_{z}(x) =-eV \frac{m\lambda}{\pi^2 v_F}
\int_{0}^{1} dy\sin({x} \sqrt{1-y}/\xi)e^{-x\sqrt{y}/\xi}
\nonumber\\
+\frac{eV}{16\pi^2 mv_F^2 x^2} \sum_\beta ~ [\sin{(2k^\beta
x)}-2k^\beta x\cos{(2k^\beta x)]} \nonumber\\
 -\frac{eV (k^{+})^2}{4\pi^2 mv_F^2} \int_0^{1} dy ~ \sin[xk^+(\sqrt{y}+
\sqrt{\delta+y})] ,
\end{eqnarray}
where $\xi^{-1}=2m\sqrt{\lambda v_F}$, and $\delta=(k^-/k^+)^2-1$.
Integrating this expression over $x$ we recover the net spin
polarization (\ref{result}).

{\it \underline{Spectral distribution of spin density}}. It is
instructive to present the results in terms of the Fourier transform
of the spin density, $s_z(q)=2\int_{0}^\infty dx s_z(x) \sin{qx}$.
From Eq.~(\ref{x}) we find $s_z(q)$ in a form of piecewise
continuous algebraic function defined in four domains. The
surprising feature of the spectral distribution revealed by this
calculation is its vanishing, $s_z(q)= 0$, in the whole
long-wavelength domain, $0<q<2k^+$. In particular, this shows the
exact cancelation between normal and evanescent modes. For larger
values of $q$ we obtain to the leading order in $\lambda$,
\begin{equation}
\label{spectr} s_z(q)=\frac{eV q}{16\pi m v_F^2} \left\{
\begin{array}{cl} -1, & 2k^+<q<2mv_F,\\ 1,& 2mv_F<q<2k^-,
\\-2/(q\xi)^{4}, & 2k^- <q.
\end{array} \right.
\end{equation}
The plot of the spectral distribution is illustrated in Fig.~3.
Remarkably, the net spin polarization (given
by $\pi^{-1}\int dq s_z(q)/ q$) comes from the large-$q$ tail
($\propto q^{-3}$) in the spectral density $s_z(q)$.

\underline{\it Conclusion}. In this paper we solved analytically a
problem of mesoscopic spin-Hall effect in a confined 2D electron
system. We presented general arguments why the net spin polarization
in a ballistic spin-Hall effect is {\it independent} of the boundary
potential and confirmed the result by a straightforward calculation
for the hard-wall boundary, for which the analytical solution was
obtained. The spectral distribution of spin density consists of two
narrow peaks of opposite sign whose heights are virtually
independent of the small spin-orbit coupling constant. Surprisingly,
long-wavelength contributions from evanescent and normal modes {\it
exactly} cancel each other. Understanding the level of universality
of this result for arbitrary boundary potentials remains a
challenging problem.

We acknowledge fruitful discussions with A. Andreev, B. Halperin, M.
Raikh and O. Starykh. The work was supported by DOE, Award
No.~DE-FG02-06ER46313 and by SFB TR
 12.
\begin{figure}
\includegraphics[width=5.0cm,angle=-90]{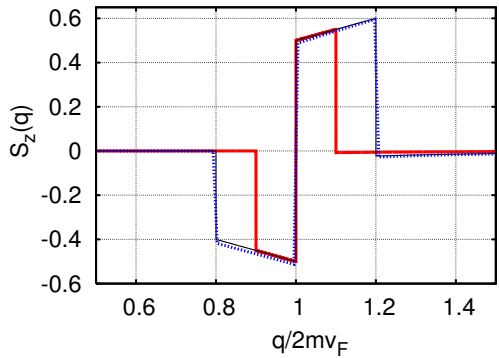}
\caption{(Color online) Spectral distribution (\ref{approximate}) of
spin density in units of $ eV/4\pi v_{F}$ for different values of
spin-orbit coupling constant. Solid (red) line: $\lambda/v_F=0.1$,
Solid (black) line: $\lambda/v_F=0.2$. Dotted (blue) line shows
Fourier transform of the exact Eq.~(\ref{x}) for $\lambda/v_F=0.2$.}
\label{fig2}
\end{figure}


\begin{thebibliography}{50}

\bibitem{DP} M.I. Dyakonov, V.I. Perel, Phys. Lett. A {\bf 35}, 459 (1971).

\bibitem{exp1} Y.K. Kato {\it et.~al}, 
Science {\bf 306}, 1910 (2004); V.
Sih {\it et.~al},
 Nature Physics {\bf 1}, 31 (2005).

\bibitem{exp2}
J. Wunderlich {\it et.~al}, 
Phys. Rev. Lett. {\bf 94}, 047204 (2005).

\bibitem{exp3} S.O. Valenzuela and M. Tinkham, Nature {\bf 442}, 176 (2006).

\bibitem{H} J.E. Hirsch, Phys. Rev. Lett. {\bf 83}, 1834 (1999).

\bibitem{Z}  Shufeng Zhang, Phys. Rev. Lett.  {\bf 85}, 393 (2000).

\bibitem{ERH} H.A. Engel, B.I. Halperin, and E.I. Rashba,  Phys. Rev. Lett.
{\bf 95}, 166605 (2005).

\bibitem{DS} W.-K. Tse, S. Das Sarma, Phys. Rev. Lett. {\bf 96}, 056601
(2006).

\bibitem{MNZ}  S. Murakami, N. Nagaosa, and S.-C. Zhang, Science {\bf 301}, 1348
(2003).

\bibitem{Sinova} J. Sinova {\it et.~al}, 
Phys. Rev. Lett. {\bf 92}, 126603 (2004).

\bibitem{reviews} S. Murakami, Adv. in Solid State
Phys. {\bf 45}, 197 (2005), cond-mat/0504353; H.A. Engel, E.I.
Rashba, and B.I. Halperin, cond-mat/0603306.

\bibitem{MSH} E. G. Mishchenko, A. V. Shytov, and B. I. Halperin,
Phys. Rev. Lett. {\bf 93}, 226602 (2004).

\bibitem{BNM}  A. A. Burkov, A. S. N\'u$\tilde {\text n}$ez, and A. H. MacDonald,
Phys. Rev. B {\bf 70}, 155308 (2004).

\bibitem{Chu} A.G. Mal'shukov {\it et.~al},  
Phys. Rev. Lett.
{\bf 95}, 146601 (2005).

\bibitem{AB} I. Adagideli and G. E. W. Bauer, Phys. Rev. Lett. {\bf 95},
256602 (2005).

\bibitem{GBD} V. M. Galitski, A. A. Burkov, and S. Das Sarma, Phys.
Rev. B {\bf 74}, 115331 (2006).

\bibitem{Blei} O. Bleibaum, Phys. Rev. B {\bf 74}, 113309 (2006).

\bibitem{Tser} Ya. Tserkovnyak {\it et.~al}, 
cond-mat/0610190.

\bibitem{RGD} R. Raimondi {\it et.~al}, 
cond-mat/0701629.

\bibitem{NWS} K. Nomura {\it et.~al}, 
Phys. Rev. B {\bf 72}, 245330 (2005).

\bibitem{nik} B.K. Nikolic {\it et.~al}, 
Phys. Rev. B {\bf 72}, 075361 (2005);  
Phys. Rev. Lett. {\bf 95}, 046601 (2005); E.M. Hankiewicz {\it
et.~al}, 
Phys. Rev. B
{\bf 70}, 241301(R) (2004).

\bibitem{Datta} S. Datta, {\it Electronic Transport in Mesoscopic
Systems}, (Cambridge University Press, 1995).

\bibitem{usaj} G. Usaj and C.A. Balseiro, Europhys. Lett. {\bf 72},
621 (2005); A. Reynoso, G. Usaj and C.A. Balseiro, Phys. Rev. B {\bf
73}, 115342 (2006).

\bibitem{SG} T.D. Stanescu and V. Galitski, Phys. Rev. B {\bf 74},
205331 (2006).

\bibitem{diff} V. Sih {\it et.~al}, 
Phys. Rev. Lett.
{\bf 97}, 096605 (2006).

\bibitem{EEL} S.I. Erlingsson {\it et.~al}, 
Phys. Stat. Sol. C {\bf 3}, 4317 (2006).

\bibitem{BR} Yu.A. Bychkov and E.I. Rashba, J. Phys. C {\bf 17} 6039
(1984).

\bibitem{Dress} G. Dresselhaus, Phys. Rev. {\bf 100}, 580 (1955).

\bibitem{both} When both Rashba and linear Dresselhaus
($\lambda_D$) couplings are present the methods developed in this
Letter are still fully applicable. In particular, the expression (1)
is valid provided that we change, $\lambda^2\to
\lambda^2-\lambda^2_D$.

\bibitem{SM} P.G. Silvestrov and E.G. Mishchenko,
Phys. Rev. B {\bf 74}, 165301 (2006).

\bibitem{ZSM} V.A. Zyuzin, P.G. Silvestrov, and
E.G. Mishchenko, in preparation.

\bibitem{evanesc} To obtain the
Green's function in the evanescent domain ($k^+<k_y<k^-$) one has to
substitute $k_x^+\to i\kappa$ into $B_+$, Eq.~(\ref{beta}), and use
the Hermitian conjugate $B^\dagger_+$ instead of $B^*_+$ in
Eq.~(\ref{psi}). In addition, vanishing of exponentially growing
terms requires the substitution $k_x^+\to i\kappa$ ($k_x^+\to
-i\kappa$) to be made in ${\hat C}_R$ (${\hat C}_A$), see
Eq.~(\ref{c}).

\end{thebibliography}
\end{document}